\documentclass[aps, prd, twocolumn, superscriptaddress, longbibliography]{revtex4-1}

\usepackage{amsmath}
\usepackage{amssymb} 
\usepackage{graphicx}

\usepackage{float}

\usepackage[caption=false]{subfig}
\usepackage{enumerate}
\usepackage{color}
\usepackage{tensor}
\usepackage{mathrsfs}
\usepackage{hyperref}
\usepackage[normalem]{ulem}
\usepackage[dvipsnames]{xcolor}
\usepackage{accents}

\begin{document}
\title{Evolution of the electric field along null rays for arbitrary observers and spacetimes}
\author{Lucas T. Santana}
\affiliation{Universidade Federal do Rio de Janeiro,
Instituto de F\'\i sica, \\
CEP 21941-972 Rio de Janeiro, RJ, Brazil}
\author{Jo\~ao C. Lobato}
\affiliation{Universidade Federal do Rio de Janeiro,
	Instituto de F\'\i sica, \\
	CEP 21941-972 Rio de Janeiro, RJ, Brazil}
\author{Isabela S. Matos}
\affiliation{Universidade Federal do Rio de Janeiro,
	Instituto de F\'\i sica, \\
	CEP 21941-972 Rio de Janeiro, RJ, Brazil}
\author{Maur\'\i cio O. Calv\~ao}
\affiliation{Universidade Federal do Rio de Janeiro,
Instituto de F\'\i sica, \\
CEP 21941-972 Rio de Janeiro, RJ, Brazil}
\author{Ribamar R. R. Reis}
\affiliation{Universidade Federal do Rio de Janeiro,
Instituto de F\'\i sica, \\
CEP 21941-972 Rio de Janeiro, RJ, Brazil}
\affiliation{Universidade Federal do Rio de Janeiro, Observat\'orio do Valongo, 
 \\CEP 20080-090 Rio de Janeiro, RJ, Brazil}

\begin{abstract}
We derive, in a manifestly covariant and electromagnetic gauge independent way, the evolution law of the electric field $E^\alpha = Ee^\alpha\ (e^\alpha e_\alpha=1)$,  relative to an arbitrary set of instantaneous observers along a null geodesic ray, for an arbitrary Lorentzian spacetime, in the geometrical optics limit of Maxwell's equations in vacuum. We show that, in general, neither the magnitude $E$ nor the direction $e^\alpha$ of the electric field (here interpreted as the observed polarization of light) are parallel transported along the ray. For an extended reference frame around the given light ray, we express the evolution of the direction $e^\alpha$ in terms of the frame's kinematics, proving thereby that its expansion never spoils parallel transport, which bears on the unbiased inference of intrinsic properties of cosmological sources, such as, for instance, the polarization field of the cosmic microwave background (CMB). As an application of the newly derived laws, we consider a simple setup for a gravitational wave (GW) interferometer, showing that, despite the (kinematic) shear induced by the GW, the change in the final interference pattern is negligible since it turns out to be of the order of the ratio of the GW and laser frequencies.
\end{abstract}

\maketitle
\section{Introduction}\label{I}
The principle that light (in vacuum) follows null geodesics is of supreme importance in relativity, astronomy and cosmology. It is legitimate to expect that it might be derived from the field equations of electrodynamics in a regime in which 
the picture of photons holds. This is precisely one of the outcomes of the geometrical 
optics (GO) limit of Maxwell's equations in curved spacetimes, which also emerges from studying the 
bi-characteristic curves of the electromagnetic field in vacuum \cite{Courant1962}. By analogy with the Wentzel-Kramers-Brillouin procedure in quantum mechanics \cite{Merzbacher1998}, 
it is common to look for asymptotic solutions of Maxwell's equations  
where the wavelength of light is much shorter than typical length scales of variation for both the metric (\textit{e.g.,}
a curvature radius) and the amplitude of the electromagnetic field \cite{Misner1973, Schneider1992, Harte2019}. 

There are at least two usual approaches to GO, differing in their choice of the fundamental field to expand: (i) either the electromagnetic potential $A^\alpha$ \cite{Misner1973, Stephani2004, Ellis2012, Straumann2013, Harte2019}, or (ii) the electromagnetic field $F^{\alpha\beta}$ \cite{Ehlers1967, Anile1989, Schneider1992, Perlick2000, Dolan2018}.
In either case, the approximation scheme assumes that the expression of the fundamental field 
in vacuum is close to that of a monochromatic homogeneous plane wave, factored into a rapidly varying phase 
and a slowly varying amplitude. In both approaches, it is easily demonstrated that (A) light rays are null geodesics, and that, in quantum language, (B) the photon number is conserved. In the potential approach, one also shows that (C) the ``polarization vector'', defined as the unit vector in the direction of $A^\alpha$ in the Lorenz gauge, is perpendicular to the rays and is parallel transported along them. This approach, however, is of course not manifestly electromagnetic gauge invariant and, in particular, the relation between the mentioned ``polarization vector'' and (the direction of) the electric field is not usually explicitly shown up.

Here, in contrast, we systematically employ the electromagnetic field approach, guaranteeing, right from the start, the  gauge independence of our results, and avoiding the introduction of auxiliary quantities that cannot be measured in an experimental setup. In particular, we consistently take the polarization vector to be the usual unit direction of the electric field \cite{Born2002, Jackson1999}.

The results derived here might be relevant whenever we want to infer or estimate intrinsic properties of an object with unknown motion by observing its light intensity and polarization. That is the case for the many probes of state-of-the-art cosmology and astrophysics, such as:  the polarization field of the cosmic microwave background (CMB), whose B modes carry information on the existence of primordial gravitational waves generated by inflation \cite{Kamionkowski2016, Ade2017}, that may be affected by gravitational lensing \cite{Marozzi2017, DiDio2019, Raghunathan2019}; the polarization of light coming from supernovae, giving clues of a possible anisotropy 
in the explosion events \cite{Wang2008, Bulla2015, Cikota2019}; the accurate determination of black hole masses by means of signatures in the polarization of light emitted by their accretion discs \cite{Afanasiev2019}; the study of standard sirens as detected by laser interferometry \cite{Abbott2017} and their optical counterpart \cite{Abbott2017b}; the description of light traveling through Sagnac interferometers \cite{Schreiber2013, Belfi2017, DiVirgilio2017}.

Our main goal is to deduce, in a manifestly covariant and electromagnetic gauge independent way, the evolution law of the electric field, Eq.~(\ref{main-result}), or, equivalently, of light intensity and polarization, Eqs. (\ref{amplitude-evolution}) and (\ref{polarization-evolution}), relative to an arbitrary set of instantaneous observers along a null geodesic, for a generic Lorentzian spacetime, in the GO limit of Maxwell's equations in vacuum. Therefrom we derive (i) the role played by the kinematics of the frame of reference on the propagation of the polarization, Eq.~(\ref{polarization-kinematics}), and (ii) apply the complete electric field evolution law to a toy gravitational wave interferometer (cf.~\ref{V}). Our signature is $(-,+,+,+)$ and the terminology regarding instantaneous observer, observer and reference frame follows \cite{Sachs1977}.
\section{Field approach to geometrical optics}\label{II}
The field approach to the geometrical optics (GO) approximation of Maxwell's equations in vacuum,
\begin{subequations}
	\label{Maxwell}
	\begin{eqnarray}
		{F^{\mu\nu}}_{;\nu} &=& 0\,, \label{Gauss} \\
		F_{[\mu\nu;\lambda]} &=& 0\,,  \label{Faraday}
	\end{eqnarray}
\end{subequations}
is established by searching for solutions of these field equations in the form of a one-parameter  
family \cite{Ehlers1967, Anile1989, Schneider1992, Perlick2000, Dolan2018}:
\begin{subequations}
	\label{ansatz}
	\begin{eqnarray}
	F_{\mu \nu}(x, \epsilon) &=& f_{\mu \nu}(x, \epsilon) e^{i S(x)/\epsilon}\,, \label{ansatz_product} \\
	f_{\mu \nu}(x, \epsilon) &:=& \sum_{n = 0}^N f_{(n)\mu \nu}(x) \epsilon^n \quad (N \ge 0)\,. 	\label{amplitude}
	\end{eqnarray}
\end{subequations}

In general, the amplitude $f_{\mu\nu}$ is a complex antisymmetric smooth tensor field  and the phase $S(x)/\epsilon$ is a real smooth scalar field, where $\epsilon$ is a dimensionless perturbation parameter proportional to the 
wavelength of the wave. This \textit{Ansatz} 
generalizes the monochromatic homogeneous plane wave solution of Maxwell's equations in Minkowski spacetime, and is expected to represent, in the limit $\epsilon 
\to 0$, a rapidly oscillating function of its phase, with a slowly varying amplitude. Moreover, the vector field 
defined by
\begin{equation} 
	\label{wave-vector}
	k_{\mu}(x) :=  S_{,\mu}(x)
\end{equation} 
is supposed to have no zeros in the considered region, and should be 
interpreted as the electromagnetic wave vector. Finally, $f^{(0)}_{\mu \nu}$ is assumed to vanish  at most in a set of measure zero. 
Inserting Eq.~(\ref{ansatz}) into Maxwell's equations (\ref{Maxwell}) and assuming that $\epsilon \to 0$, 
the two leading order relations imply the following equations:
\begin{subequations}
	\label{algebraic-constraints}
	\begin{eqnarray}
	k_\nu F^{\mu \nu} & = 0\,, \label{faraday-eigen} \\
	k_{[\lambda} F_{\mu \nu]} & = 0\,, \label{hodge-eigen}
	\end{eqnarray}
\end{subequations}
\begin{equation}
	\label{null-tangent}
	k_{\mu} k^{\mu} = 0\,,
\end{equation}
and
\begin{equation}
\label{faraday-transport-equation}
F_{\mu \nu;\lambda}k^{\lambda} + \frac{1}{2} {k^{\lambda}}_{;\lambda} F_{\mu \nu} = 0\,.
\end{equation}
To derive these equations we assume, here and henceforth, that $f^{(0)}_{\mu \nu}$ is a good approximation 
for the amplitude of the electromagnetic field, \textit{i.e.}, $F_{\mu \nu} (x, \epsilon) \approx f_{(0)\mu \nu}(x) e^{i S(x)/\epsilon}$. 
\section{Electric field transport law}\label{III}
We start by choosing some light ray of the stream of photons, and introducing an instantaneous 
observer field along such a ray $u^{\mu}(\vartheta)$ ($u_{\mu} u^{\mu} = -1$), where $\vartheta$ is an affine 
parameter of the null geodesic. The electric field seen by $u^{\mu}$ is $E_{\mu} := F_{\mu \nu} u^{\nu}$, which we also split in its magnitude $E := \left(E_{\mu} \bar{E}^{\mu}\right)^{1/2}$ 
(the bar denotes complex conjugation) and corresponding (complex) polarization $e_{\mu} := E_{\mu}/E$. 
From Eq.~(\ref{hodge-eigen}) and the antisymmetry of $F_{\mu\nu}$, it is easy to derive
\begin{equation}
\label{field-decomposition}
	F_{\mu \nu} = \frac{(k_{\mu} E_{\nu} - k_{\nu} E_{\mu})}{\omega_e}= \frac{E}{\omega_e}(k_{\mu} e_{\nu} - k_{\nu} e_{\mu})\,,
\end{equation}
where $\omega_e := -k_{\mu} u^{\mu}$ is the frequency attributed to the electromagnetic wave 
by $u^\mu$. From Eq.~(\ref{field-decomposition}), one can show that the scalar field $E/\omega_e$ is an instantaneous observer-independent quantity.

Now, Eq.~(\ref{faraday-transport-equation}) implies a transport law for the 
electric field along a null geodesic:
\begin{equation}
	\label{main-result}
	\frac{DE^{\mu}}{d\vartheta} + \frac{1}{2} {k^{\lambda}}_{;\lambda} E^{\mu} = \left( \frac{k^{\mu} E^{\nu} - k^{\nu} E^{\mu}}{\omega_e}\right) \frac{Du_{\nu}}{d\vartheta}\,,
\end{equation}
where $D/d\vartheta$ is the absolute derivative along the curve. This is our 
key result (valid for the magnetic field as well), from which several other important consequences will emerge. Eq.~(\ref{main-result}) is readily 
decomposed into transport equations for both the scalar $E/\omega_e$ 
and the polarization vector, namely:
\begin{subequations}
\begin{eqnarray}
	\label{amplitude-evolution}
	\frac{d}{d\vartheta} \left( \frac{E}{\omega_e} \right) + \frac{1}{2} {k^{\lambda}}_{;\lambda} \left(\frac{E}{\omega_e} \right) = 0\,, \\
	\label{polarization-evolution}
	\frac{De^{\mu}}{d\vartheta} = k^{\mu} \left(\frac{e^{\nu}}{\omega_e} \frac{Du_{\nu}}{d\vartheta} \right)\,.
\end{eqnarray}	
\end{subequations}
Equation~(\ref{amplitude-evolution})  leads to the evolution of the specific photon number density. By defining the corresponding intensity 
measured by $u^{\mu}$ as $I := g_{\mu \nu} \,\textrm{Re}(E^{\mu})\,\textrm{Re}( E^{\nu})$, and
using the transport equation for the area of the light beam's cross section \cite{Ellis2012}, Eq.~(\ref{amplitude-evolution}) 
gives the (instantaneous) observer-independent conservation of photon number \cite{Rindler1991}, thus demonstrating law (B).

Since $k^\alpha = \omega_e(u^\alpha + n^\alpha)$, 
where $-n^\alpha$ is the line of sight relative to $u^\alpha$, 
Eq.~(\ref{polarization-evolution}) also shows that the purely spatial part of the evolution 
of polarization is entirely along $n^\mu$ itself, so that 
\begin{equation}
\label{screen_transport}
s_{\alpha\beta}\,\dfrac{De^\beta}{d\vartheta}=0\,,
\end{equation}
where $s_{\alpha\beta}:=g_{\alpha\beta}+u_\alpha u_\beta-n_\alpha n_\beta$ is the screen projection tensor (cf.~\cite{Dehnen1973, Fleury2015a}); we might phrase this as stating that the direction of the electric field is ``screen transported'' along the light rays, not parallel transported. Moreover, using the Leibniz rule on Eq.~(\ref{screen_transport}), one arrives again at Eq.~(\ref{polarization-evolution}), attesting their equivalence (cf.~\ref{VI}).
\section{Kinematic quantities and the analogy with redshift}\label{IV}
\begin{figure}
	\includegraphics[scale=0.3]{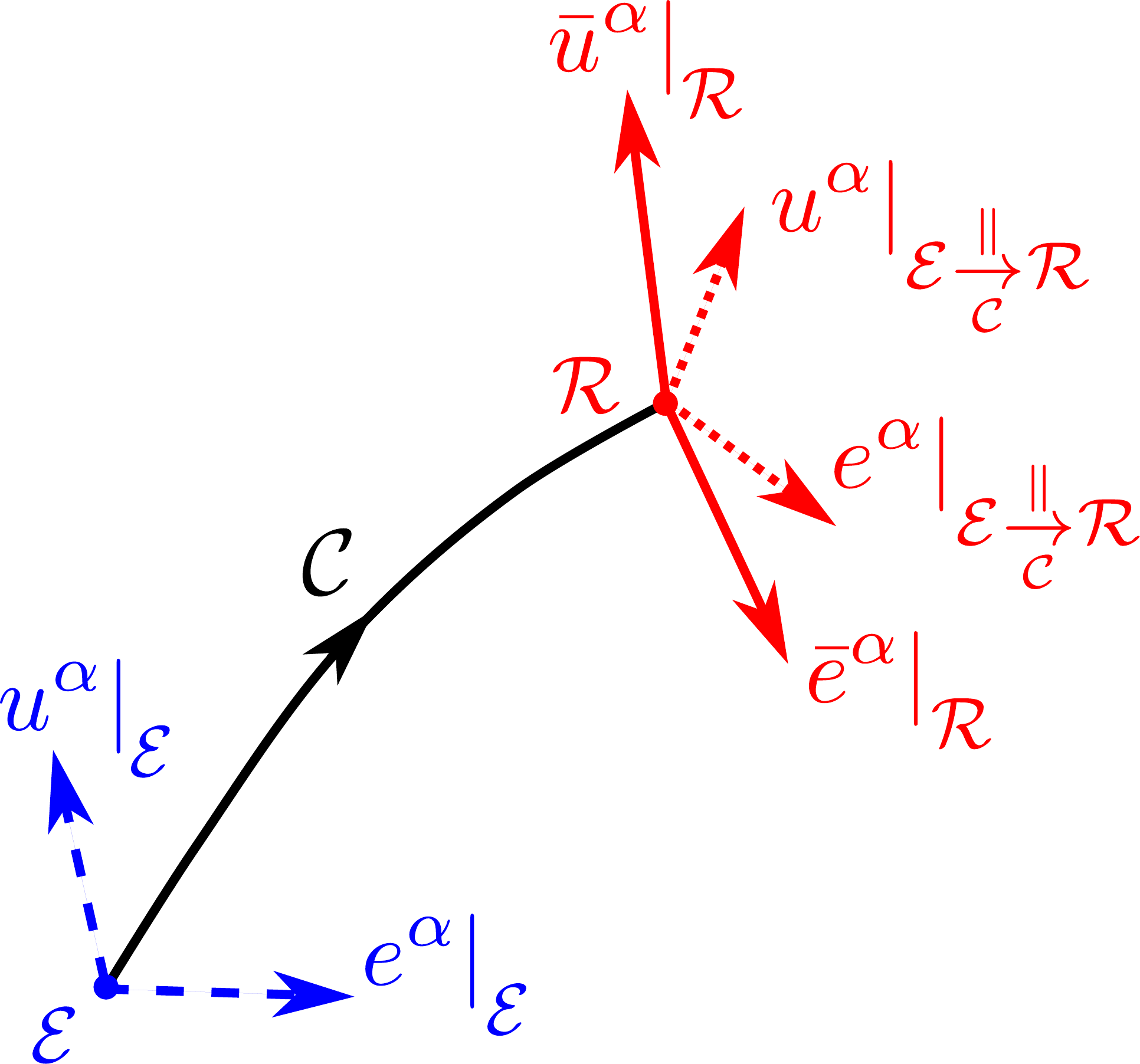}\caption{Parallel transport of (dashed, blue) $u^\alpha|_\mathcal{E}$ ($e^\alpha|_{\mathcal E}$), at event ${\cal E}$, to (dotted, red) $u^\alpha|_{{\mathcal E}\underset{\mathcal C}{\overset{\parallel}{\rightarrow}}{\mathcal R}}$ ($e^\alpha|_{{\mathcal E}\underset{\mathcal C}{\overset{\parallel}{\rightarrow}}{\mathcal R}}$), at event ${\cal R}$, along (solid, black) null geodesic ray ${\cal C}$, and then local boost, at ${\cal R}$, from $u^\alpha|_{{\mathcal E}\underset{\mathcal C}{\overset{\parallel}{\rightarrow}}{\mathcal R}}$  ($e^\alpha|_{{\mathcal E}\underset{\mathcal C}{\overset{\parallel}{\rightarrow}}{\mathcal R}}$) to (solid, red) $\bar{u}^\alpha|_{\mathcal R}$  ($\bar{e}^\alpha|_{\mathcal R}$).}
	\label{fig:transport}
\end{figure}

From a purely theoretical point of view, in Eq.~(\ref{polarization-evolution}), one sure can always choose  $u^\alpha(\vartheta)$ to be the parallel transport of an 
instantaneous observer $u^\alpha|_{\mathcal E}$ at some given initial event $\mathcal E$ (cf. Fig.~\ref{fig:transport}), which implies that
the polarization vector is parallel transported as well. However, at the most relevant events, the emission ${\mathcal E}$ and the reception ${\mathcal R}$, there might be some preferred instantaneous observers defined by other more practical or physical impositions, such as, \textit{e.g.}, isometries or the real motion of the detector, $\bar{u}^\alpha|_{\mathcal R}$, at ${\mathcal R}$ (or of the source, at ${\mathcal E}$, for that matter), which, in general, would entail $\bar{u}^\alpha|_{\mathcal R}\neq u^\alpha|_{{\mathcal E}\underset{\mathcal C}{\overset{\parallel}\rightarrow}{\mathcal R}}$. Of course, if we do make such a choice, we will have to carry out a local boost at the reception (or emission) event, in order to translate our calculation of $e^\alpha(\vartheta)$ to the effectively observed polarization, $\bar{e}^\alpha|_{\mathcal R}$. This is analogous to what happens to the 
redshift between two instantaneous observers at different events. In fact, one can always 
think of both frequency and polarization shifts as local effects, due to a boost between the actual instantaneous observer and the parallel transported one \cite{Synge1960, Narlikar1994} (cf.~\ref{VI}). 
Nevertheless, from an experimental perspective, it is convenient to consider 
two arbitrary instantaneous observers at different events and to decompose, in particular, the redshift between 
them, depending on the situation, as gravitational, cosmological, 
etc \cite{Ehlers1961, *Ehlers1993, Ellis2012}. If these two 
instantaneous observers belong to a reference frame $u^{\mu}(x)$ whose gradient is
\begin{equation}
\label{kinematic-quantities}
u_{\mu ; \nu} = -a_{\mu} u_{\nu} + \frac{1}{3} \Theta h_{\mu \nu} + \sigma_{\mu \nu} + \Omega_{\mu \nu} \,,
\end{equation}
where $h_{\mu\nu}:=g_{\mu\nu}+u_\mu u_\nu$ is the rest space projection tensor, and $a_{\mu}$, $\Theta$, $\sigma_{\mu \nu}$ and $\Omega_{\mu \nu}$ are the kinematic 
quantities of the reference frame, respectively, its acceleration, expansion, shear, and 
vorticity \cite{Ehlers1961, *Ehlers1993, Ellis2012}, this redshift decomposition
follows immediately from \cite{Ehlers1961, *Ehlers1993, Ellis2012}
\begin{equation}
	\label{redshift-evolution}
	\frac{1}{\omega_e}\frac{d\omega_e}{d\vartheta} = - \left(\frac{1}{3}{\Theta} + a_{\mu} n^{\mu} + \sigma_{\mu \nu} n^{\mu} n^{\nu} \right) \omega_e\,.
\end{equation}
By analogy, Eq.~(\ref{polarization-evolution}) implies a similar 
effect for the polarization:
\begin{equation}
	\label{polarization-kinematics}
	\frac{De^{\mu}}{d\vartheta} = \left(a_{\nu} + \sigma_{\nu \lambda} n^{\lambda} + \Omega_{\nu \lambda}  n^{\lambda}\right) e^{\nu}k^{\mu}\,.
\end{equation}
Notice that the expansion of the reference frame never contributes to the change in 
the polarization of an electromagnetic wave  (in the GO 
regime) along any of its light rays, whereas for the redshift, it is the 
vorticity which plays no role. In addition to the purely expanding case, the polarization vector 
will be parallel transported: 
(i) if shear and vorticity vanish, 
and the acceleration is orthogonal to the electric field (\textit{e.g.}, radial light 
rays seen by static observers in Schwarzschild spacetime); (ii) if acceleration 
and vorticity vanish, and either the polarization or the line of sight is an 
eigenvector of the shear (cf.~\ref{V}); (iii) if acceleration and shear vanish, and 
the vorticity vector is orthogonal to the magnetic field.  These examples illustrate the fact that parallel 
transporting the corresponding reference frame along the null geodesic is only a sufficient condition for the
polarization to be also parallel transported. In fact, due to Eq.~(\ref{polarization-evolution}), the parallel transport of the polarization is equivalent to $e^\alpha Du_\alpha/d\vartheta=0$. Notice thus that, analogously to the redshift, there is a local boost freedom to maintain this property (cf.~\ref{VI}).

\section{Application: Gravitational wave interferometry}\label{V}
Now, we apply Eq.~(\ref{main-result}) to calculate the interference pattern in a toy 
gravitational wave (GW) interferometer. It is common to describe the effect in free particles 
caused by GW in Minkowski background using the transverse-traceless (TT) gauge 
\cite{Misner1973,Maggiore2007}. When discussing the GW influence on the output signal of an interferometric experiment, disregarding the optical expansion effect as we also do here, one often relates it exclusively with the difference $\Delta t$ of the travel times through each arm. This is commonly done in either of two ways: imposing that the electric field propagates in the $\eta_{\mu \nu}$ background metric, relative to the unperturbed inertial frame \cite{Maggiore2007} or by means of the potential vector \cite{Thorne2017}. Since the TT comoving frame experiences a 
shearing motion when the wave passes by, implying a nonzero $ u_{\mu ; \nu}$ 
(cf.~Eq.~(\ref{kinematic-quantities})) {and the Christoffel symbols in ($\ref{main-result}$) are nonzero in the TT coordinates, it is reasonable to expect that taking into 
account the full propagation deduced in this work might lead to possible corrections to the final intensity beyond the usual phase shift contribution. 
 
The simpler interferometer configuration whose arms are aligned with the  shear eigenvectors is not a good first choice if we want to assess, for example, the relevance of
Eq.~(\ref{polarization-evolution}) in the measured interference pattern, since we verified that
in such configuration, the polarization vector field, though no longer seen by a parallel 
transported reference frame, is still parallel along the corresponding null geodesics (cf.~\ref{IV}). Figure 
\ref{fig:interferometers} shows the usual Michelson-Morley setup. 
\begin{figure}[h]
	\subfloat[]{\includegraphics[width=0.26\textwidth]{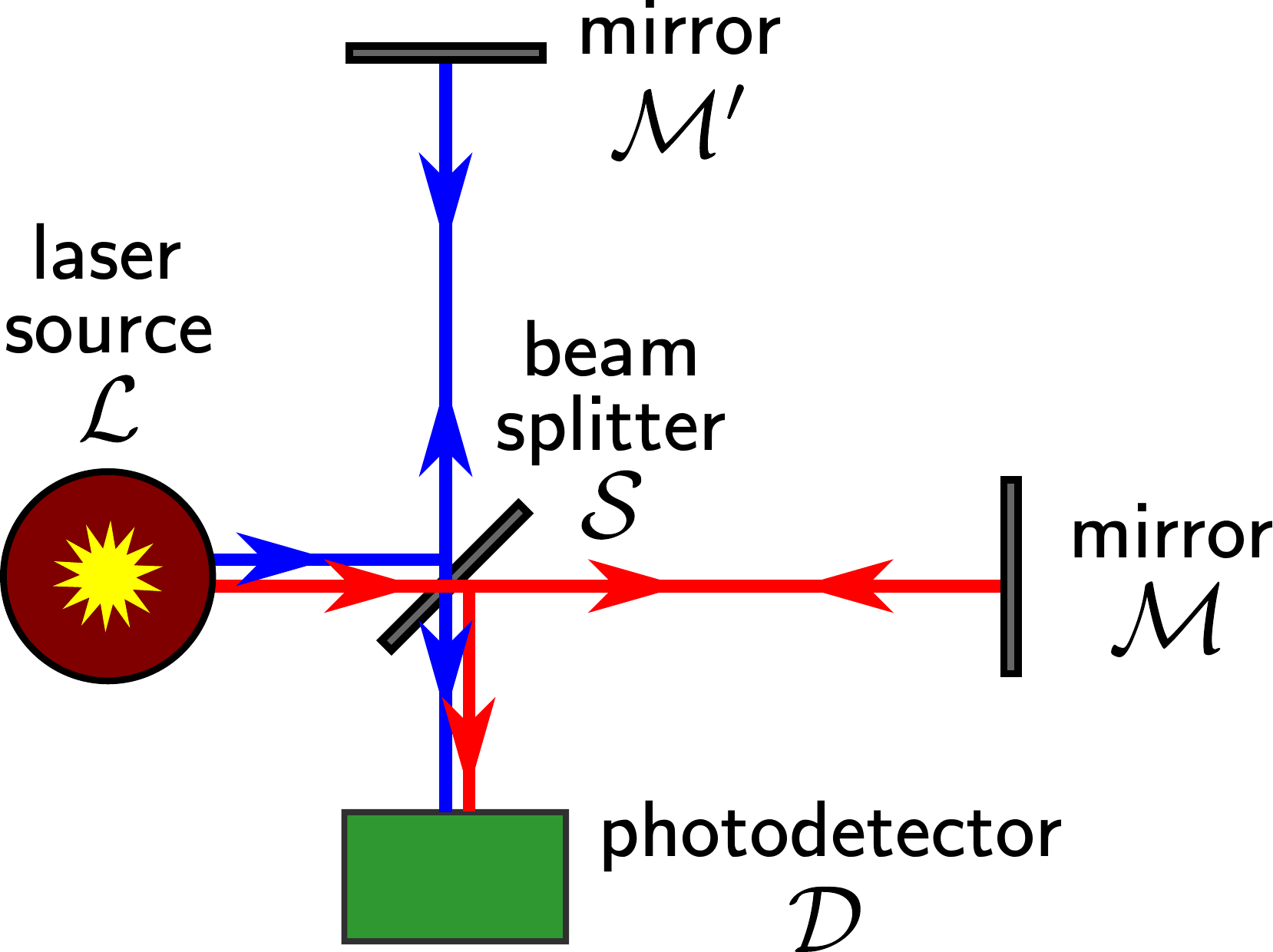}}\hfill
	\subfloat[]{\includegraphics[width=0.2\textwidth]{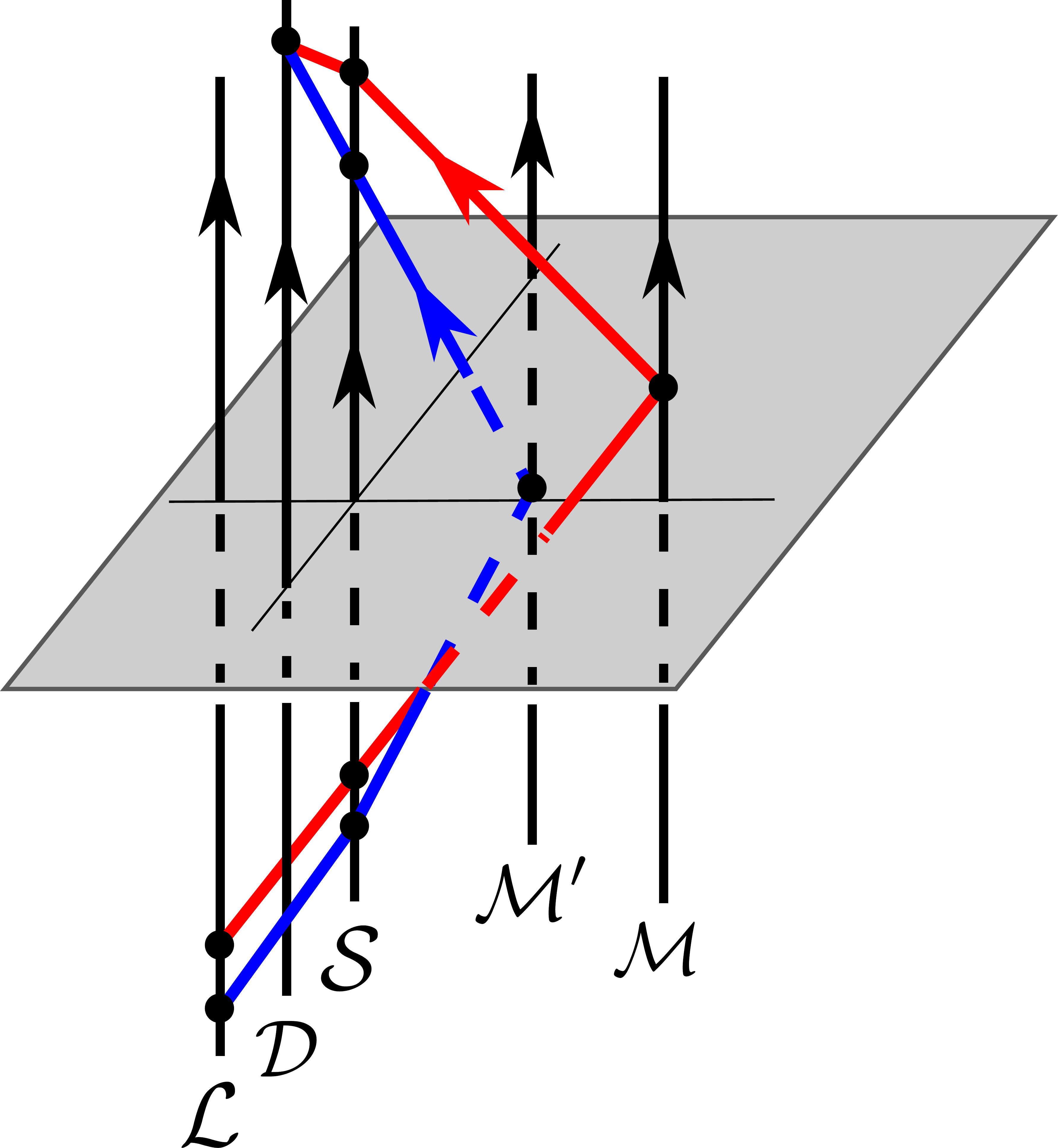}}
	\caption{(a) Spatial diagram of a Michelson-Morley GW interferometer. (b) Corresponding spacetime diagram, with relevant worldlines: tilted ones for the null geodesic arcs and vertical ones, for the several devices.}
	\label{fig:interferometers}
\end{figure}
We choose plane wave packets for each GW polarization propagating perpendicularly to the apparatus:
\begin{align}
ds^2 =& -dt^2 + dx^2 + [1 - h_+(t - x)]dy^2 \nonumber \\ 
&+ [1 + h_+(t-x)]dz^2 - 2h_{\times}(t-x) dy dz\,.
\end{align}

We allow for distinct lengths of the arms and all calculations are performed up to first 
order in GW amplitudes. From the general evolution equation  \cite{Ellis2012} for the optical expansion, $\hat{\Theta}:={k^\alpha}_{;\alpha}$,  it follows that a vanishing initial value implies, in Ricci-flat spacetimes, $\hat{\Theta}= 0$ along the whole  curve. This is valid because the electromagnetic 
field has vanishing optical shear
to leading order in the GO approximation \cite{Harte2019}. For simplicity, this initial condition is imposed here.

Using Eq.~(\ref{main-result}), we determine the electric field 
along the null geodesic arcs within the interferometer. At the end of the round trips (from ${\cal S}$ to either ${\cal M}$ or ${\cal M}'$ and back to ${\cal S}$), we sum the propagated fields in each arm obtaining the total real electric field $E_T^{\mu}$. We then calculate the complete intensity $I_c := g_{\mu \nu} E^{\mu}_{T} E^{\nu}_{T}$, in which one notices a correction to the standard $I_{\Delta t}$ associated with just the plain difference in travel times, calculated through $dE^{\mu}/d \vartheta = 0$ \cite{Maggiore2007}. These two quantities can be expressed as a sum of the Minkowski spacetime contribution ($I_M$, corresponding to $h_+ = h_{\times} = 0$) and the one due to the presence of the GW, so that we define:
\begin{equation}
\delta I_{c, \Delta t} := I_{c, \Delta t} - I_M\,.
\end{equation}   
As expected, $\delta I_{\Delta t}$ is the contribution due to the fact that, under the presence of the GW, there is a slight extra induced difference between the light travel times in each arm. This term is also present in $\delta I_c$, together with an additional effect associated with small redshifts (or blueshifts) acquired by light. The change in frequency can be thought as a consequence of the  relative velocities between the extremities of the arms inherent on the frame's kinematic shear motion. An interesting feature of $\delta I_c$ is that the RHS of Eq.~(\ref{polarization-evolution}) happens not to alter it, and thus one could assume the polarization vector to be parallel transported without any harm to the interference pattern prediction in this configuration.

We assess the relevance of the
new found propagation law, Eq.~(\ref{main-result}), in this situation, by computing the order of magnitude of the discrepancy, which turns out to be, for each GW mode of frequency $\omega_g$,
\begin{equation}\label{discrepancy}
\frac{\delta I_c - \delta I_{\Delta t}}{\delta I_{\Delta t}} \sim \frac{\omega_g}{\omega_e|_{{}_{\vartheta = 0}}} < 10^{-10}
\end{equation}
for all current and future planned detectors, such as LIGO \cite{Abbott2016} 
and LISA \cite{Danzmann2017}, implying that this kinematic contribution can be 
safely disregarded in the theoretical description of these experiments \cite{Thorne2017}.
These derivations will be presented in detail in a future work.
\section{Discussion}\label{VI}
In this work, as key results of GO approximation, we showed, for general spacetimes 
and arbitrary instantaneous observers, that the ratio $E/\omega_e$ and the instantaneous observer-dependent polarization $e^\alpha$ evolve according to  Eqs.~(\ref{amplitude-evolution}) and (\ref{polarization-evolution}), respectively. These two equations combined are equivalent to Eq.~(\ref{main-result}). We also derived the evolution of polarization in terms of the kinematics 
of the corresponding frame of reference [cf.~Eq.~(\ref{polarization-kinematics})] and established 
an analogy with redshift. As an important result, 
Eq.~(\ref{polarization-kinematics}) shows that in a purely expanding reference frame 
($a_{\mu} = \sigma_{\mu \nu} =  \Omega_{\mu \nu}=0, \Theta \neq 0$) the polarization  is indeed parallel transported. This is precisely the case of the Hubble flow in an exact Friedmann-Lema\^itre-Robertson-Walker universe, which thus ensures that, in this context, to derive intrinsic polarization properties of (unperturbed) cosmological observables, it is safe to use parallel transport with respect to this background frame; the extension to perturbed models should be pursued in further works.

We then applied Eq.~(\ref{main-result}) to assess systematic effects due to the 
kinematics of the reference frame comoving with a GW interferometer (cf.~Eq.~(\ref{discrepancy})).
The dominant contribution to $\delta I_{\Delta t}$ and $\delta I_{c}$ arises from the difference between light travel times in the arms
via $\Delta t \, \omega_e|_{{}_{\vartheta = 0}}$. 
For our particular setup, the kinematics of the adapted frame does not affect the final 
intensity pattern appreciably. This is a pleasant result, particularly so in view of the 
recent interest in GW related parameter estimation \cite{Abbott2017}.
It is important to note that when the ratio $\omega_g/\omega_e|_{{}_{\vartheta = 0}}$ 
is close to unity, the frequency shift and travel time contributions to the intensity 
become comparable, and one should not, in principle, neglect the former. However, this condition represents the transition between the geometrical
and wave optics descriptions of light, so that 
Eqs.~(\ref{faraday-eigen}), (\ref{hodge-eigen}), and (\ref{faraday-transport-equation}) 
are no longer guaranteed.

Dehnen \cite{Dehnen1973} had already obtained analogues to our Eqs.~(\ref{polarization-evolution}) and (\ref{screen_transport}) during 
his study of the gravitational Faraday effect  for null electromagnetic fields. The main 
difference between our approach and his is that we considered GO, 
representing an approximate solution of Maxwell's equations, whereas he studied exact null fields. We also supposed 
the vector $k^{\mu}$ to be the gradient of the wave's phase in Eq.~(\ref{ansatz}), 
whilst he assumes that $k^{\mu}$ may possess nonvanishing 
curl. Na\"{i}vely, our result seems to be 
less general than Dehnen's. However, the optical shear of an exact null electromagnetic
field is precisely zero \cite{Robinson1961}, while a GO one can, in principle,
correspond to a light bundle that may be sheared by a gravitational lens \cite{Harte2019}.

We would also like to make two brief comments on the potential approach as described in the \textit{Introduction}: (i) the plane wave Ansatz in two different gauges might not give rise to the same set of derived physical electromagnetic fields; (ii) even so, however, in the Lorenz gauge, it is possible to show that $E^{\mu} = i\omega_e {s^{\mu}}_{\nu}A^{\nu}$ and then in fact obtain Eq.~(\ref{polarization-evolution}) from the parallel transport law of the unit ``polarization vector'' along $A^{\mu}$.

Equations~(\ref{field-decomposition}) and (\ref{polarization-evolution}) allow us to clear up some subtle issues. First, at any event on the null ray, we can perform a Lorentz boost,
\begin{equation}
\bar{u}^\alpha=\gamma(u^\alpha + vc^\alpha)\,,   \label{Lorentz-boost}
\end{equation}	
where $vc^\alpha$ is the relative (3-)velocity between the instantaneous observers ($c^\alpha u_\alpha=0, c^\alpha c_\alpha = 1$), and derive, from Eq.~(\ref{field-decomposition}), the transformation law for the polarization vector:
\begin{equation}
\bar{e}^\alpha = e^\alpha + \dfrac{v e^{\mu} c_{\mu}}{\omega_e(1-vn^\beta c_\beta)}\,k^\alpha\,.
\label{transformation-polarization}
\end{equation}	
Thus, the \textit{polarization plane}, defined as the 2-dimensional tangent subspace spanned by $k^\alpha$ and $e^\alpha$, is instantaneous observer-independent. 

Now let us consider no longer purely local results (cf. Fig.~\ref{fig:transport}), regarding Eq.~(\ref{polarization-evolution}) as a first-order ODE system for $e^\alpha(\vartheta)$, with $u^\alpha(\vartheta)$ smoothly prescribed \textit{a priori}, of course. First, together with Eq.~(\ref{transformation-polarization}), it implies that the plane of polarization is parallel transported along the ray \cite{Perlick1993}. Second, alternatively to parallel transporting $u^{\alpha}$ and then performing a boost at a final event as previously pointed out, if one decides to choose $u^{\alpha}$ without appealing to any local Lorentz boost whatsoever, with $u^{\alpha}|_{\mathcal E}$ and $u^\alpha|_{\mathcal R}$ independently given, this will constrain the set of allowed smooth instantaneous observers $u^\alpha(\vartheta)$ such that the RHS of Eq.~(\ref{polarization-evolution}) in general will not vanish any more. Finally, Eq.~(\ref{transformation-polarization}) again shows that, starting with a field $u^{\alpha}$ such that $e^{\mu}(\vartheta)$ is parallel transported, another field for which this property still holds, once the same initial condition for the polarization is assumed, is degenerate up to boosts satisfying $e^{\beta} c_{\beta} = 0$.

An important consequence of the equivalence between Eqs.~(\ref{polarization-evolution}) and (\ref{screen_transport}) for the usual description of polarization properties of the CMB \cite{Fleury2015a, DiDio2019} is that no influence due to the kinematics of the frame resulting from perturbing  the background Hubble flow of a cosmological model is lost if one solely uses the screen space projected equation.

As for the future, we plan to consider more realistic interferometer setups, by generalizing the motion of the mirrors to arbitrary time-like geodesics, allowing for  
GWs to arrive at oblique angles, and investigating the influence of the optical expansion term on Eq.~(\ref{main-result}).  We shall also look 
for an alternative feasible Earth-based experiment that could detect the effect of the acceleration of Schwarzschild static observers on nonradially propagating light, 
analogous to the one used to measure the gravitational redshift in the M\"ossbauer effect \cite{Pound1964}.
\begin{acknowledgments}
The authors would like to thank Professors Ruth Durrer and Phillip Bull for valuable 
discussions and suggestions to the work.
L.T.S. and I.S.M. thank Brazilian funding agency CNPq for PhD scholarships GD 141325/2017-8 and GD 140324/2018-6, respectively. 
J.C.L. thanks Brazilian funding agencies CAPES and FAPERJ for MSc scholarships 31001017002-M0 and 2016.00763-4, respectively.
\end{acknowledgments}
%
%

\end{document}